\documentclass[twocolumn,showpacs]{revtex4}
\usepackage{dcolumn}
\usepackage{amsmath}
\usepackage[xdvi]{graphicx}

\newcommand{\bra}{\left\langle}
\newcommand{\ket}{\right\rangle}

\newcommand{\ca}{{\rm C}_{\alpha}}
\newcommand{\vecr}{{\boldsymbol r}}
\newcommand{\vecv}{{\boldsymbol v}}

\newcommand{\vecxi}{{\boldsymbol \xi}}

\begin{document}

\title{Violation of the fluctuation-dissipation  
theorem in a protein system 
}
\author{Kumiko Hayashi and Mitsunori Takano} 
\affiliation
{Department of Physics, Waseda University, Tokyo 169-8555, Japan}

\date{\today}

\begin{abstract}
We report the results of  molecular dynamics simulations of 
the protein myosin  carried out with an elastic network model.  
Quenching the system, we observe  glassy behavior of a density 
correlation function and a density response function that are 
often investigated in structure glasses and spin glasses.  
In the equilibrium, the fluctuation-response relation,  
a representative relation of the fluctuation-dissipation theorem, 
holds that the ratio of the density correlation function to 
the density response function is equal to the temperature of 
the environment. We show that in the  quenched system that we 
study, this relation can be violated.  In the case that this 
relation does not hold, this ratio can be regarded as an 
effective temperature.  We  find that this  effective temperature 
of myosin  is higher than the temperature of the environment.  
We discuss the relation between this effective temperature and 
energy transduction  that occurs after ATP hydrolysis in the 
myosin molecule.  
\end{abstract}

\pacs{87.14.Ee, 05.70.Ln, 05.40.-a}
\maketitle

\section{Introduction}

\begin{figure}
\begin{center}
\includegraphics[width=1.8in]{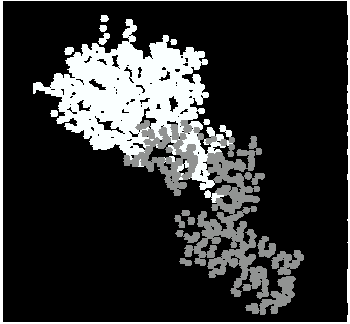}
\caption{
Schematic depiction of the myosin molecule (1KK7 \cite{1kk7}). 
The horizontal axis 
denotes the $x$ direction, and the vertical axis denotes the $y$ direction.  
According to our convention,  
the $\alpha$-carbon atoms, $\ca$ with  with $i=1,\cdots,770$   
belong to the head substructure (white),  
and those  with $i=771,\cdots,1101$  
belong to the tail substructure (gray).   See the next section for 
the way how to number $\ca$ atoms. 
}
\label{myosin}
\end{center}
\end{figure}

Proteins possess  complex structures and, consequently, complex motion.   
Such complexity might be necessary to  carry out  the functions exhibited 
by living organisms. Statistical properties of the types of complex 
structure and motion characterizing protein molecules have been studied 
using various approaches, in particular, that employing energy 
landscapes  \cite{frau}.  Recently, with a computation of the density 
of states for a  G\=o-like model of a protein, statistical properties 
were  investigated  using the idea of inherent structures \cite{nakagawa}.  
An inherent structure is a subset of the configuration space that 
represents the local minima of an energy landscape, originally proposed 
to study the dynamics of liquids \cite{still}.  Although the 
idea of  inherent structures is very important to understand protein 
dynamics, it is difficult to directly experimentally investigate  the 
inherent structures of a protein.   For this reason, it would be useful 
if we could characterize the  energy landscape  of a protein in forms 
of experimentally measurable quantities, such as a density correlation 
function or a density response function.

In structure glasses, which, like protein molecules, also possess 
energy landscapes with many local minima, statistical properties 
are often studied by computing the density correlation function and 
the density response function \cite{ritort,berthier,g1,g2,g3,g4}.  
In equilibrium, these quantities satisfy the fluctuation-response 
relation, a representative relation in the fluctuation-dissipation  
theorem \cite{hs}.    
This relation means the equality of the ratio of the density 
correlation function to the density response function and the 
temperature of the environment \cite{hs}. For glassy systems,  which 
exhibit slow relaxation and are inherently non-equilibrium, it has 
been reported that the  fluctuation-response relation is violated  
\cite{ritort,berthier,g1,g2,g3,g4}; that is, the ratio of the relevant 
density correlation function and density response function 
is not equal to the temperature of the environment. The slow 
relaxation displayed by a glassy system, which results from  the 
nature of its energy landscape, with many local minima,  can be 
characterized by a quantity representing the degree of violation 
of  the fluctuation-response relation.  In some cases that the 
fluctuation-response relation is violated, the ratio of the density 
correlation function to the density response function has interpreted 
as an ``effective temperature,''  and there are studies addressing 
the question of whether this effective temperature can play the role 
of the temperature in non-equilibrium systems 
\cite{ritort,berthier,g1,g2,g3,g4}.

In this paper, we report the results of simulations of molecular 
dynamics employing an elastic network model of sub-fragment 1 of 
a myosin molecule, which  is composed of a head substructure, with 
ATP-binding and actin-interacting sites, and a  tail substructure,  
with a long alpha-helix bound by two light chains (see Fig. \ref{myosin}). 
With the hypothesis that appropriate density correlation function and 
density response function, which have not yet been fully exploited 
in studies of protein dynamics, can be used  to characterize the 
glassy behavior of  protein molecules,  we compute such quantities  
in the case of a quenched system.  While elastic network models 
have been used to study the elastic properties of equilibrium 
fluctuations \cite{zheng,nav},  such models have not been used to study 
non-equilibrium behavior.  In this paper, we show that glassy, 
non-equilibrium behavior can also be described with this class of models.    

Because the effective temperature that we employ represents the degree to 
which the behavior of the  system is ``glassy,''  we quantitatively 
investigate the complexity of the myosin's head through use of this 
effective temperature.  More specifically,   we compute the values of the   
effective temperature for both the head and the tail, and we compare these 
values with the temperature of the environment.  We also seek to 
elucidate the origin of the observed glassy behavior through   
analysis that investigates the inherent structures \cite{still}. 
 We find that the inherent structure for the model we study has 
a resemblance to the structure of a structural isomer of myosin 
determined by X-ray crystallography.

\section{Elastic network model}

In our study, a protein molecule is regarded as consisting of 
$\alpha$-carbon  
atoms, $\ca$, employing a coarse-grained representation of 
amino acid residues \cite{ela1,ela2,ela3}.   A $\ca$ atom is a 
constituent of the carbon skeleton of a protein molecule.  

The position and velocity of the $i$-th $\ca$ in a myosin molecule  
are denoted by $\vecr_i=(x_i,y_i,z_i)$ and 
$\vecv_i=(v_{xi},v_{yi},v_{zi})$, where $i=1,\cdots,N$,  
and $N$ is the total number of  $\ca$ atoms  in the myosin molecule.  
 Here,   $\ca$ atoms are numbered in the order of 
the heavy chain, the essential light chain and the regulatory light 
chain. 

The Hamiltonian of our model is given by 
\begin{equation}
H(\{\vecr_i\},\{\vecv_i\})=\sum_{i=1}^N \frac{m}{2}|\vecv_i|^2 
+V(\{\vecr_i\})+V_{\rm trap}(\{\vecr_i\}),
\label{hami} 
\end{equation}
where $m$ is the mass of the $i$-th $\ca$, and  $V(\{\vecr_i\})$ 
is the interaction potential 
\begin{eqnarray}
V(\{\vecr_i\})&=&\sum_{i}^{N-1} \frac{k_1}{2}
(|\vecr_i-\vecr_{i+1}|-|\vecr_i^0-\vecr_{i+1}^0|)^2   \nonumber \\
&+& \sum_{i}^{N-1} \frac{k_2}{2}
(|\vecr_i-\vecr_{i+2}|-|\vecr_i^0-\vecr_{i+2}^0|)^2  \nonumber \\
&+&  \sum_{i}^{N-1} \sum_{j=i+3}^{N} \frac{k_3}{2}
(|\vecr_i-\vecr_j|-|\vecr_i^0-\vecr_j^0|)^2 .   
\label{gopote}
\end{eqnarray} 
We stipulate that 
$k_{\ell}=0$ ($\ell=1,2,3$) if $|\vecr_i^0-\vecr_{j}^0| 
> r_{\rm c}$, where $r_{\rm c}$ is a cut-off length, and 
we set $k_2=0.5 k_1$ and $k_3=0.1 k_1$.  With this form of 
$V(\{\vecr_i\})$, the configuration of 
the native structure, $\{\vecr_i^0\}$,   is most stable.  
The values of the parameters $\vecr_i^0$ used here were taken from the 
RCSB Protein Data Bank \cite{pdb}.  Here,  we chose the values 
of $\vecr_i^0$ so as to obtain the structure of the myosin 
molecule 1KK7 \cite{1kk7}.  Also, these values are such that 
the center of mass  is at $(0,0,0)$ and  the unit of length is 
\AA.  

The function 
$V_{\rm trap}(\{\vecr_i\})$ that appears in (\ref{hami}) is a 
trapping potential that plays the role of 
 an optical potential, acting  to fix the molecule.   
This potential is given by  
\begin{equation}
V_{\rm trap}(\{\vecr_i\}) 
=\sum_{i=350}^{400} \frac{1}{2}|\vecr_i-\vecr_i^0|^2
+\sum_{i=800}^{850} \frac{1}{2}|\vecr_i-\vecr_i^0|^2.
\label{trap} 
\end{equation}

The time evolution of the system is described by the 
Langevin equation ($i=1,\cdots,N$)   
\begin{eqnarray}
m\frac{{\rm d}\vecv_i}{{\rm d}t}&=&-\gamma\vecv_i
-\frac{\partial H}{\partial \vecr_i}+\vecxi_i(t), \\ 
\frac{{\rm d}\vecr_i}{{\rm d}t}&=& \vecv_i, 
\label{lan}
\end{eqnarray}
where $\vecxi_i=(\xi_{xi},\xi_{yi},\xi_{zi})$ is Gaussian white noise 
that satisfies 
\begin{equation}
\langle 
\xi_{\alpha i}(t) \xi_{\beta j}(t')\rangle 
 =2\gamma T \delta(t-t')
\delta_{\alpha,\beta}\delta_{i,j}.   
\end{equation}
Here, 
$T$ is the temperature of the environment (with the Boltzmann constant  
set to $1$), and $\gamma$ is the friction constant of the solvent.  
Here, $\bra \ \ket$ denotes the average over all noise histories. 
We set the parameters used in our numerical simulations as 
$m=1$, $\gamma=0.01$, $k_1=1$, $N=1101$ and $r_{\rm c}=10$.

\begin{figure}
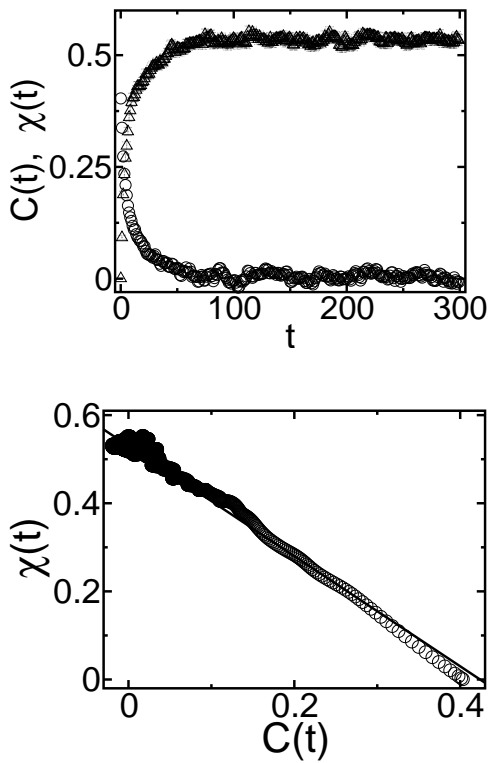

\begin{center}
\includegraphics[width=2.5in]{echic.eps}
\end{center}
\begin{center}
\vspace{3mm}
\includegraphics[width=2.5in]{echic2.eps}
\end{center}
\caption{ ({\it Top})
$C(t)$ (circles) and $\chi(t)$ (triangles) 
plotted as  functions of time in the equilibrium case with $T=0.8$, 
$\Delta=0.05$ and $k=2\pi/10$.   
({\it Bottom}) $\chi(t)$ as a function of 
$C(t)$ in the case $T=0.8$.  The slope of the line here is 
$-1/T$. In both graphs, the data points are obtained by 
averaging $2000$ independent trajectories, and 
the statistical error bars are smaller than the symbols. 
}
\label{figechic2}
\end{figure}

\section{
Equilibrium behavior: Fluctuation-response relation at high 
temperature
}

In equilibrium, the ratio of the density correlation function to the  
density response function is equal to the temperature of the 
environment.  This relation among the density correlation function, 
the density response function and the temperature of the environment 
is called the fluctuation-response relation  
 (For a review of the fluctuation-dissipation theorem, including  the 
fluctuation-response relation, see Ref. \cite{hs}).

In order to examine whether the fluctuation-response relation holds 
for our system, we first introduce a density 
response function. In this section,    we consider the relaxation 
process that  results in the case that the system is initially 
in equilibrium, and then at $t=0$ the perturbing potential   
\begin{equation}
\sum_{i=1}^N V_{\rm p}(y_i)\equiv \sum_{i=1}^N 
\Delta \cos(k y_i) 
\label{pote}
\end{equation}   
is added to the Hamiltonian (\ref{hami}).   Note  that we consider 
a perturbation along the $y$ direction, which is approximately 
parallel to the long axis of the myosin (see Fig. \ref{myosin}). 
We have also studied the cases with perturbations along different  
directions, and the results obtained in all cases are qualitatively 
the same as those obtained along the $y$ direction. 
In this relaxation process, the susceptibility, $\chi(t)$,  is 
defined as   
\begin{equation}
\chi(t)\equiv -\frac{\bra \hat{\rho}(k,t) - \hat{\rho}(k,0)   
\ket^{V_{\rm p}}}{\Delta},
\end{equation} 
where $\bra \ \ket^{V_{\rm p}}$ denotes the statistical average under 
this relaxation process.   
Writing the density in the $y$ direction as  
$\rho(y,t)\equiv \sum_{i=1}^N \delta(y-y_i(t))/N$, we denote  
its Fourier transform by $\hat{\rho}(k,t)$:  
\begin{eqnarray}
\hat{\rho}(k,t)&\equiv&\int_{-\infty}^{\infty}
{\rm d}y \rho(y,t)\cos(ky)  \nonumber \\ 
&=&\frac{1}{N}\sum_{i=1}^N \cos(k y_i(t)). 
\end{eqnarray} 

Next, we define the density correlation function, which is computed 
in equilibrium without adding $V_{\rm p}$,  as  
\begin{equation}
C(t)\equiv \bra \hat{\rho}(k,t)\hat{\rho}(k,0) \ket N. 
\end{equation}
Then, the following is one representation of the fluctuation-response 
relation:  
\begin{equation}
R(t)=-\frac{1}{T}\frac{{\rm d} C(t)}{{\rm d}t}, 
\label{frr}
\end{equation}
where $t\ge 0$ \cite{hs}. Here, the  density response function, 
$R(t)$, defined as the time  derivative of $\chi(t)$:   
\begin{equation}
R(t)\equiv \frac{{\rm d} \chi(t)}{{\rm d} t}.  
\end{equation} 
In numerical experiments, we computed $\chi(t)$ in order to investigate 
the behavior of $R(t)$. 
 
In the upper graph of Fig. \ref{figechic2},  $\chi(t)$ and $C(t)$ are 
plotted as functions of time with $T=0.8$, $k=2\pi/10$ and 
$\Delta=0.05$. In the lower graph of Fig. \ref{figechic2},  $\chi(t)$ 
is plotted as a function of $C(t)$.  The fact that the slope of the 
lower graph is $-1/T$  indicates that the fluctuation-response 
relation (\ref{frr}) holds in equilibrium (the high temperature 
regime). Note that we set  $k=2\pi/10$ here because the lengths of 
the alpha helices in myosin are of the order of $10$ \AA.   Glassy 
behavior of the density correlation  and the density response  is 
observed in the case $k=2\pi/\ell$ with $\ell \ge 10$, as described 
below.

\section{Non-equilibrium behavior I: Glassy phenomenon 
 at low temperature
}

In this section, we consider simulations in which we quenched the 
system from $T=0.5$ to $T=0.05$ in order to investigate its glassy 
behavior.  In this model, room temperature roughly corresponds to 
$T=0.1$, which is estimated by comparing the mean square 
fluctuations of the $\ca$ atoms in the model with those obtained 
in an all-atom model (data not shown).  
Therefore, the quenched temperature $T=0.05$ 
roughly corresponds to $150$ K, which is below the glass transition 
temperature of a real protein \cite{austin}.  Note that myosin 
remains in a folded form even under conditions  corresponding to $T=0.5$.

In our simulations, we quench the system at $t=-t_{\rm w}$, and then 
we begin investigating the behavior of the system at $t=0$. 
The quantity $t_{\rm w}$ is called the ``waiting time.''   
 In general, relaxational glassy systems do not possess 
invariance with respect to time translation,  and both correlations and 
responses decay more slowly as the age of the system increases.  
Thus, in such systems, correlation functions and response functions 
depend on $t_{\rm w}$.

\begin{figure}
\begin{center}
\includegraphics[width=2.5in]{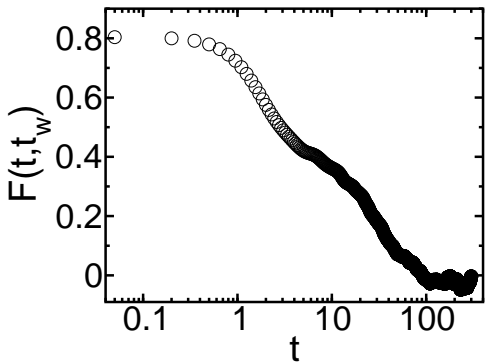}
\end{center}
\caption{$F(t,t_{\rm w})$ as a function of time for  the case 
$t_{\rm w}=100$.  The data points are obtained by 
averaging $2000$ independent trajectories, and 
the statistical error bars are smaller than the 
symbols. 
}
\label{figauto}
\end{figure}

In order to elucidate the glassy behavior of our system, 
we first introduce the auto-correlation function 
\begin{equation}
F(t,t_{\rm w})\equiv \bra 
\frac{1}{N}\sum_{i=1}^N [\cos(y_i(t)-y_i(0))-\cos(y_i^0-y_i(0))]
\ket_{t_{\rm w}}, 
\label{auto}
\end{equation}
where $\bra \ \ket_{t_{\rm w}}$ represents the statistical 
average under the relaxation we consider. (Auto-correlation functions 
are often employed  in the study of glassy systems 
\cite{ritort,berthier,g1,g2,g3,g4}.)  The quantity $F(t,t_{\rm w})$  
is a measure of the similarity of  the configuration of the system 
at time $t$  to that at time $0$; note that the quantity  
$\cos\left(y_i(t)-y_i(0)\right)$ is equal to $1$ at $t=0$ and 
decreases as $t$ increases.  The second term on the righthand 
side of (\ref{auto}) ensures that  $F(t,t_{\rm w})=0$ for  large $t$,  
because the configuration of the system, $\{\vecr_i(t)\}$,  
fluctuates about that of the native structure, $\{\vecr_i^0\}$,  at 
large $t$.  Computing $F(t,t_{\rm w})$, we can¡¡elucidate the slow 
relaxation of the configuration of the system.  In Fig. \ref{figauto},  
we plot $F(t,t_{\rm w})$ as a function of time for the case 
$t_{\rm w}=100$.  The form of the decay, which is not exponential, 
reveals the slow relaxation of the configuration. 

\section{Non-equilibrium behavior II: Violation of the 
fluctuation-response relation at low temperature
}

In an aging system, the susceptibility, density response function 
and density correlation function, which depend on $t_{\rm w}$,  are 
defined by 
\begin{eqnarray} 
&&
\chi(t,t_{\rm w})\equiv -\frac{ 
\bra \hat{\rho}(k,t)-\hat{\rho}(k,0)\ket_{t_{\rm w}}^{V_{\rm p}} 
}{\Delta},  
\label{gres}\\
&&
R(t,t_{\rm w})\equiv \frac{{\rm d} \chi(t,t_{\rm w})  }{{\rm d} t}, \\ 
&&
C(t,t_{\rm w})\equiv\bra \hat{\rho}(k,t)\hat{\rho}(k,0) \ket_{t_{\rm w}}N.  
\label{gcorre} 
\end{eqnarray}
Note that $\bra \ \ket_{t_{\rm w}}^{V_{\rm p}}$ denotes the statistical average 
under the condition that we quench the system at $t=-t_{\rm w}$, and 
switch on the perturbation potential given by (\ref{pote}) at $t=0$. 
In the upper graph of Fig. 5, $C(t,t_{\rm w})$ and 
$\chi(t,t_{\rm w})$ are plotted as  functions of time for the case 
$t_{\rm w}=100$.  The  relaxations of  $C(t,t_{\rm w})$ and 
$\chi(t,t_{\rm w})$  seem to be slow in comparison with those 
of $C(t)$ and $\chi(t)$ found in the equilibrium case with 
$T=0.8$ (see the upper graph in Fig. 2).

\begin{figure}
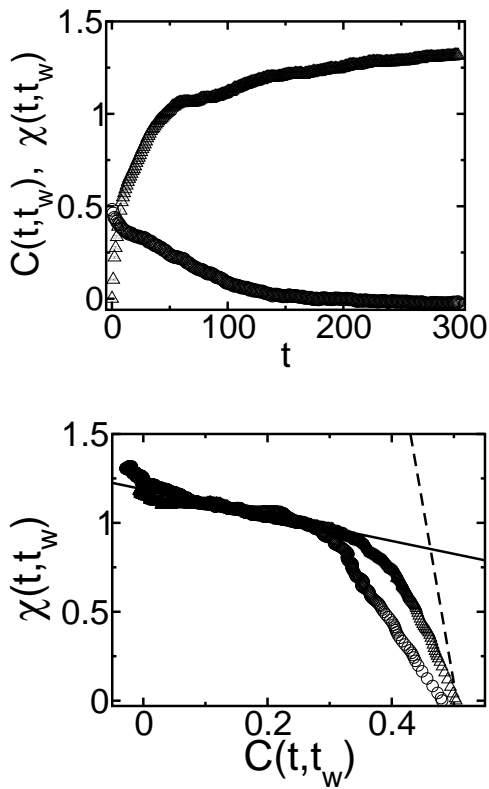

\begin{center}
\includegraphics[width=2.5in]{chic.eps}
\end{center}
\begin{center}
\vspace{3mm}
\includegraphics[width=2.5in]{chic2.eps}
\end{center}
\caption{({\it Top}) $C(t,t_{\rm w})$ (circles) and 
$\chi(t,t_{\rm w})$ (triangles)  plotted as functions of time  
in the case of an aging system, with $t_{\rm w}=100$, and $\Delta=0.05$ 
and $k=2\pi/10$.  ({\it Bottom}) $\chi(t,t_{\rm w})$ as a function of 
$C(t,t_{\rm w})$ for the cases $t_{\rm w}=100$ (circles) and 
$t_{\rm w}=200$ (triangles).  The slope of the dotted line is $-1/T$, 
where $T=0.05$. 
The slope of the solid line is  $-1/T_{\rm eff}$, where 
$T_{\rm eff}=1.3$.  In both graphs, the data points are obtained by 
averaging $2000$ independent trajectories, and 
the statistical error bars are smaller than the symbols. 
}
\label{figchic2}
\end{figure}

In the lower graph of Fig. \ref{figchic2},  we plot  
$\chi(t,t_{\rm w})$  as a function of $C(t,t_{\rm w})$ for the 
cases $t_{\rm w}=100$  and  $t_{\rm w}=200$.  It is seen that there 
are two slopes,  characterizing two time regimes,  unlike in 
the equilibrium case (see the lower graph in Fig. \ref{figechic2}). 
The results plotted in Fig. \ref{figechic2} suggest that  in the 
early regime,  the form of $\chi(t,t_{\rm w})$  as a function of 
$C(t,t_{\rm w})$ approaches a line of the slop $-1/T$, where $T=0.05$,  
as $t_{\rm w}$ increases.  (See  Fig. 16 of Ref. \cite{g4} for such 
observation of similar behavior.)  In the late time regime, the form 
of this function is again a line, but in this case, the slope differs 
significantly from $-1/T$. (Also note that in this case, similar 
linear behavior is observed for both $t_{\rm w}=100$ and $t_{\rm w}=200$.) 
Such  behavior of $\chi(t,t_{\rm w})$ and $C(t,t_{\rm w})$ is often 
observed in aging systems,  including structure glasses \cite{ritort,berthier} 
and spin glasses \cite{ritort,g1}.  In an aging system, the slope of  
$\chi(t,t_{\rm w})$  as a function of $C(t,t_{\rm w})$ in the late 
regime is $-1/T_{\rm eff}$ where  $T_{\rm eff}$ is called the effective 
temperature, and is defined through the relation 
\begin{equation}
R(t,t_{\rm w})=-\frac{1}{T_{\rm eff}}\frac{{\rm d} 
C(t,t_{\rm w})}{{\rm d}t}. 
\label{gfrr}
\end{equation}
We interpret this behavior as indicating that in the early regime, 
the system relaxes toward the local equilibrium represented by 
a meta-stable  state, while  in the late regime, the system wanders 
among many meta-stable  states, as it evolves toward the minimum of 
the free energy of the system.  These two kinds of relaxational 
processes exhibited by an aging system  are characterized by the 
two slopes $-1/T$ and  $-1/T_{\rm eff}$.  

The fact that  $T_{\rm eff}\ne T$ represents a violation of the 
fluctuation-response relation in this aging system. The lack of  
invariance with respect to time translation and the violation of 
the fluctuation-response relation are a basic properties of 
relaxational glassy systems.  The effective temperature $T_{\rm eff}$  
denotes the difference of the system from equilibrium. 

Before ending this section, we note an important point regarding 
$T_{\rm eff}$.  The inequality $T_{\rm eff}> T$ is  observed in many 
glassy systems \cite{ritort,berthier,g1,g2,g3,g4}. However, because 
there are counterexamples, the universality of this inequality has 
not yet been established.  Here, we found  $T_{\rm eff}> T$ in the 
elastic network model.  Possible implications of this inequality 
in biological contexts are discussed below. 

\section{$T_{\rm eff}$  for the head and tail}

One of the biggest differences between a structure glass and 
a protein is that a protein has characteristic substructures that 
are believed to cause a smoothing of the energy landscape, which makes the  
collective motion necessary for biological functions possible.  As 
mentioned above, a myosin molecule is composed of a head,  
with ATP-binding and actin-interacting sites, and a tail,   
with a long alpha-helix bound by two light chains.  
It is  interesting to compare these two parts with respect to 
$T_{\rm eff}$.  For this purpose, we   
investigated the density response function and density correlation 
function for the head  ($i=1,\cdots,770$) and  
the tail ($i=771,\cdots,N$), and computed $T_{\rm eff}$ for each.

We define  the Fourier transform of the density for the head and tail 
as 
\begin{eqnarray}
\hat{\rho}_{\rm h}(k,t)&\equiv& \frac{1}{770}\sum_{i=1}^{770}\cos(k y_i), \\
\hat{\rho}_{\rm t}(k,t)&\equiv& \frac{1}{N-770}\sum_{i=771}^{N}\cos
(k y_i). 
\end{eqnarray}
Then, we use $\hat{\rho}_{\rm h}(k,t)$ and $\hat{\rho}_{\rm t}(k,t)$ 
instead of $\hat{\rho}(k,t)$ in (\ref{gres}) and (\ref{gcorre}).  
Also, letting $N_{\rm h}=770$ and $N_{\rm t}=N-770$ be the total numbers  
of  $\ca$ atoms  in the head and tail, respectively, we use these 
in place of  $N$ in (\ref{gres}) and (\ref{gcorre}).   We note here 
that when computing the response functions,  we add a perturbing  
potential of the form given in (\ref{pote}) only to the substructure 
under investigation.

\begin{figure}
\begin{center}
\includegraphics[width=2.5in]{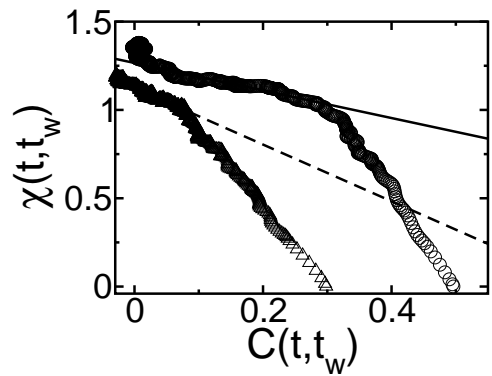}
\end{center}
\caption{ $\chi(t,t_{\rm w})$ as a function of $C(t,t_{\rm w})$ 
in the case $t_{\rm w}=100$ and $k=2\pi/10$ for the head (circles) 
and the tail (triangles) of the myosin molecule.  The slope of the 
solid line is equal to $-1/1.3$, while   
the slope of the dotted line is equal to $-1/0.6$. 
 The data points are obtained by 
averaging $2000$ independent trajectories, and  
the statistical error bars are smaller than the symbols. 
}
\label{fightchic}
\end{figure}

In Fig. \ref{fightchic}, we plot $\chi(t,t_{\rm w})$ as a function 
of $C(t,t_{\rm w})$ for both the head and the tail, respectively.  
We find that the effective temperature, $T_{\rm eff}$, for the head 
is higher  than that for the tail.  This implies 
that the structure of the head is more glassy than that of the tail.

\section{Discussion}

In conclusion, 
we investigated the glassy behavior of an elastic network model of the 
myosin molecule  by studying    
the density correlation function and the density response function 
in the case that we  quench  the system from $T=0.5$ to $T=0.05$.  
The glassy behavior is displayed in Figs. \ref{figchic2} and 
\ref{fightchic},  where the susceptibility,  $\chi(t,t_{\rm w})$, 
is plotted as a function of the correlation function, $C(t,t_{\rm w})$. 
We found that $T_{\rm eff}$ defined in (\ref{gfrr}) was not equal to $T$, 
and thus that the fluctuation-response relation is  violated.  
 We also compare the degrees of the violation  of the 
fluctuation-response relation for both the head  and  tail  
substructures of myosin, individually, and we found that 
 $T_{\rm eff}$ is higher for  the head  than for the tail. 
In the following, we discuss two points related to these main results. 

\subsection{Effective temperature of the myosin molecule}

Although further studies are required to ascertain physically clear 
interpretations of the effective temperature, the fact that $T_{\rm eff}$ 
is higher  than $T$ is intriguing,  because it may be related to energy 
transduction taking place after ATP hydrolysis in a myosin molecule.  

An acto-myosin system, of which a myosin molecule is  a component,  
is  a representative motor-protein system, and it  has been  studied 
in single-molecule experiments \cite{yana,yana2,yana3}.  
Upon binding of ATP to a myosin molecule, the myosin molecule 
detaches from an actin filament. ATP hydrolysis takes place in the 
detached myosin molecule. The products of the ATP hydrolysis (ADP 
and an inorganic phosphate) are then released from the myosin 
molecule, which are thought to be coupled with the force exertion 
and with the re-attachment of myosin to the actin filament. 
Recently, a single-molecule 
experiment  on an acto-myosin system  demonstrated that the force 
exertion of the myosin molecule sometimes occurs {\it after} the release of 
the bound ADP \cite{yana}.  This fact leads us to believe 
that the energy provided by the   ATP hydrolysis might be stored in 
the  myosin molecule for a short time before the force is exerted.  
If this is indeed the case,  it is important to determine the form 
in which this  energy is stored.   The effective temperature,  
$T_{\rm eff}$,  might help us solve this problem.  Because ATP hydrolysis 
takes place only in the head,  and because  $T_{\rm eff}$  is higher 
in the head than in the tail, it is  reasonable to conjecture that  
this energy storage has a close  connection with the inequality 
$T_{\rm eff} >  T$. To elucidate this connection is a future problem.

\subsection{Inherent structures}

\begin{figure}
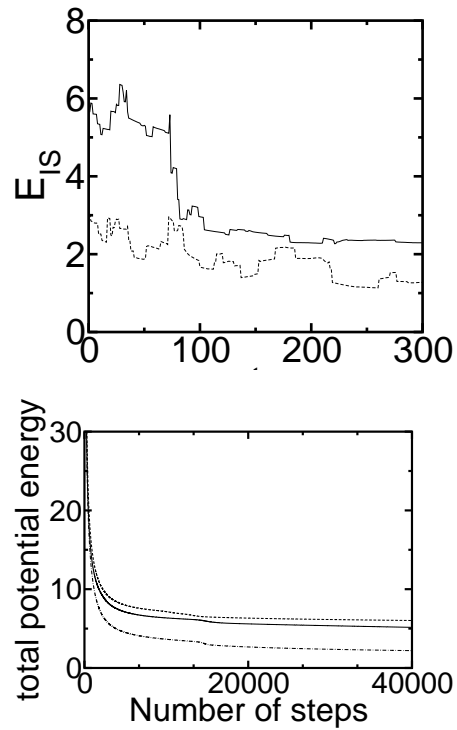

\begin{center}
\includegraphics[width=2.3in]{is.eps} 
\\
\includegraphics[width=2.3in]{is2.eps} 
\end{center}
\caption{({\it Top}) 
Two examples of $E_{\rm IS}$, computed using the steepest-descent 
 energy-minimization  
method, plotted as functions of  time in the case $t_{\rm w}=100$. 
({\it Bottom})  Three examples of the steepest-descent 
energy-minimization trajectories  are shown. Note that the three 
energy-minimization trajectories, with  different initial configuratio
ns (i.e., different ``instantaneous structures'' on  the MD trajectory), 
tend to converge at different energy values.  
}
\label{figis}
\end{figure}

Next, we discuss the origin of the glassy behavior observed in the elastic 
network model we study.  In Fig. \ref{figis}, we plot 
$E_{\rm IS}$, which is the total potential energy, $V+V_{\rm trap}$, 
 when the system is at a local minimum in the energy landscape, 
as a function of time. The quantity $E_{\rm IS}$ is computed  
by using the steepest-descent energy-minimization 
method, as has been employed in Ref.  \cite{still},  
in the case $t_{\rm w}=100$.  
From the rather discretized forms of the trajectories, it is seen that the 
system moves from one meta-stable state to another.   
We stress that the elastic network model described by (\ref{gopote})   
is not a harmonic system, and hence can possess local energy minima, 
although  the model may appear to be a harmonic system (Actually, existense 
of local energy minima in an elastic network model is indicated 
in Ref. \cite{togashi}).

\begin{figure*}
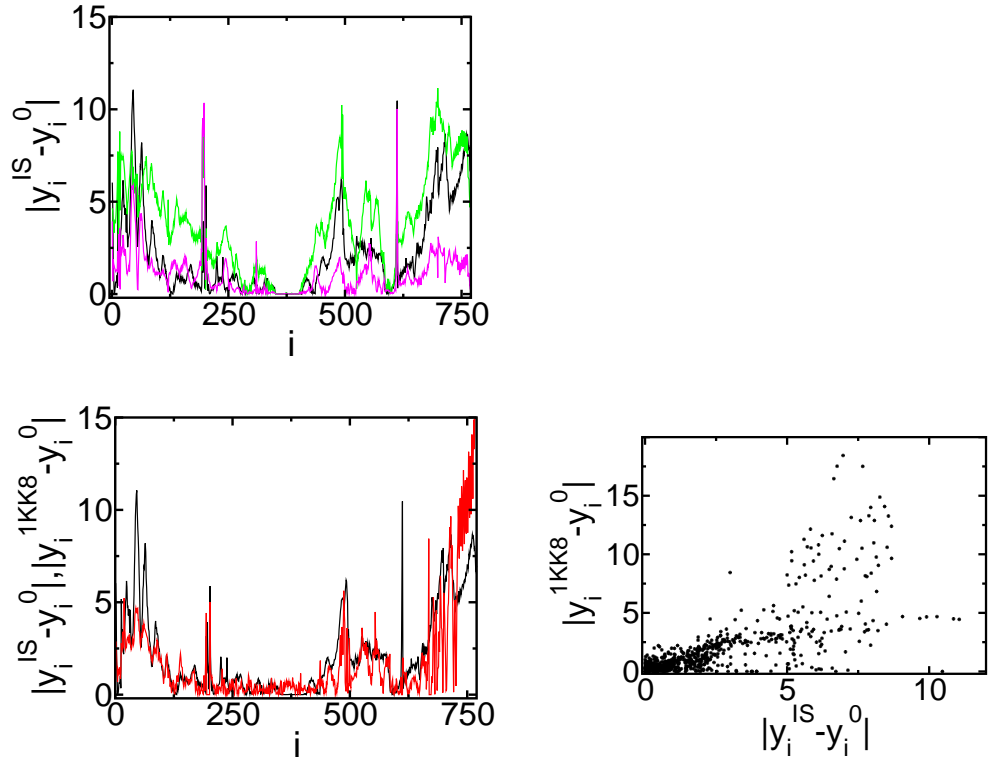

\begin{center}
\includegraphics[width=2.5in]{quasi.eps}
\hspace{6.5cm}
{ }
\\ 
\vspace{6mm}
\includegraphics[width=2.5in]{meta.eps}
\hspace{5mm}
\vspace{5mm}
\includegraphics[width=2.3in]{meta2.eps}
\end{center}
\caption{({\it Top}) Three examples (black, pink, green) of 
$|y_i^{\rm IS}-y_i^0|$ plotted as functions of $i$  
($1\le i\le 770$ , the head substructure in our definition), where 
$\{y_i^{\rm IS}\}$ is a configuration computed  using the steepest-descent 
 energy-minimization  
method, and  $\{y_i^{0}\}$ is the configuration of the structure 1KK7. 
Note that the $\alpha$-carbon atoms,   
$\ca$,  with $i=350,\cdots,400$  
are trapped by $V_{\rm trap}$ (see (\ref{trap})).    
({\it Bottom, Left})  An example of 
$|y_i^{\rm IS}-y_i^0|$ (black)  and $|y_i^{\rm 1KK8}-y_i^0|$ (red)  
plotted as  functions of $i$. Here, 
$\{y_i^{\rm 1KK8}\}$ is the configuration of the structure 1KK8.  
({\it Bottom, Right})   $|y_i^{\rm 1KK8}-y_i^0|$  plotted as a function of 
$|y_i^{\rm IS}-y_i^0|$.  
}
\label{figquasi}
\end{figure*}

It would be interesting to examine the difference between the native 
structure and the structures corresponding to the local energy minima 
(i.e., inherent structures).   The upper graph of   
Fig. \ref{figquasi}, displays 
the difference  $|y_{i}^{\rm IS}-y_{i}^0|$ between the positions 
of the $\ca$ atoms in the inherent structure and in the structure, 1KK7, 
where 
$\{y_i^{\rm IS}\}$ is the $y$ value of the position of the $i$-th 
$\ca$ atom in the  inherent structure.   
From this graph, it is seen that 
the differences for some of the $\ca$ atoms are as much as 10 \AA  
and that those $\ca$ which exhibit such large displacements are not 
localized but, rather, clustered.

Several structural isomers of  myosin 
have been experimentally found by using X-ray crystallography 
\cite{iso}.  The structure we studied here, 1KK7, 
is one with no nucleotide bound.  Therefore, it would be interesting 
to compare the inherent structures with those of myosin isomers.  
As an example, we compare one of these inherent structures with the 
structure of the 
myosin isomer with an ATP-analog bound, 1KK8 \cite{1kk7}.  
The lower graph in Fig.  \ref{figquasi} 
shows that there is a resemblance between the inherent structure 
 and the structure of 
the myosin isomer, with relatively large displacements seen near the 
N-terminal domain, the lower-50k domain, and the converter domain. 
This suggests that the elastic network model  contain information 
concerning the structures of other isomers.   Using a normal mode 
analysis for an elastic network model of a myosin molecule, Zhen 
and Doniach have shown that a structural isomer is located along 
the directions of some of the slowest modes of the structure  \cite{zheng}.   
Our results appear to be consistent  with their results.  Furthermore,  
 our results indicate the meta-stability of the isomer.   
We believe that  our results  may lead  to an extension of the    
applicability of the elastic network model of proteins,  
noting that  it has been shown that this model  can even be applied  
to the study of protein folding \cite{fold} and the investigation 
of nonlinear relaxation dynamics \cite{togashi}.   
To elucidate the range of applicability of the 
 elastic network model, we need to systematically 
study its meta-stable states   for many kinds of proteins.

\begin{acknowledgments}
The authors acknowledge K. Komori for discussions concerning 
the experimental 
studies of acto-myosin systems   
and H. Takagi for facilitating these discussions.  
We also acknowledge M. Otsuki and K. Hukushima for discussions 
of glassy systems and A. S. Mikhailov and Y. Togashi for discussions 
of elastic network 
models.  This work was supported by grants from JSPS Research 
Fellowships for Young Scientists  and the Ministry 
of Education, Science, Support and Culture of Japan.  
\end{acknowledgments}

\begin{thebibliography}{20}

\bibitem[*]{email}
Electronic address:
hayashi@jiro.c.u-tokyo.ac.jp,\\ 
mtkn@waseda.jp 

\bibitem{frau}
Frauenfelder, H., Wolynes, P. G. and Austin, R. H.  (1999) 
{\it Rev. Mod. Phys.} {\bf 71} S419-S430. 

\bibitem{nakagawa}
Nakagawa, N. and Payrard, M. (2006) {\it Proc. Natl. Acad. Sci. USA}  
{\bf 103}, 5279-5284.

\bibitem{still}
Stillinger, F. H. and  Weber, T. A. (1982) {\it Phys. Rev. A} 
{\bf 25}, 978-989. 

\bibitem{ritort}
Crisanti, A. and Ritort, F.  (2003) {\it J. Phys. A} {\bf 36}, R181-R290. 

\bibitem{berthier}
Berthier, L.  and Barrat, J-L.  (2002) {\it Phys. Rev. Lett.} {\bf 89}, 
095702. 

\bibitem{g1}
Cugliandolo, L. F., Kurchan, J., and Peliti, L.  (1997) {\it Phys. Lev. E}   
{\bf 55}, 3898-3914.  

\bibitem{g2}
Ono, I. K.,  O'Hern, C. S.,  Durian, D. J.,  Langer, S. A.,   
Liu, A. J. and Nagel, S. R. (2002) {\it Phys. Rev. Lett.} {\bf 89}, 
095703.  

\bibitem{g3}
Kolton, A. B., Exartier, R.,Culiangolo, L. F., Dom\'iguez, D. and 
Gr$\o$nbech-Jensen, N. (2002) {\it Phys. Rev. Lett.} {\bf 89}, 095703.  

\bibitem{g4}
Cugliandolo, L. F. (2002) {\it e-print} cond-mat/0210312. 

\bibitem{hs}
Hayashi, K., and Sasa, S. (2006) {\it Physica A} {\bf 370}, 407-429. 


\bibitem{zheng}
Zheng, W. and Doniach, S. (2003) {\it Proc. Natl. Acad. Sci. USA} 
{\bf 100}, 13253-135258. 

\bibitem{nav}
 Navizet, I., Lavery, R. and  Jernigan, R. L. (2004)  
{\it Proteins} {\bf 54}, 384-393. 


\bibitem{ela1}
Tirion, M. M.  (1996) {\it Phys. Rev. Lett.} {\bf 77}, 1905-1908. 
 
\bibitem{ela2}
Atilgan, A. R., Durell, S. R., Jernigan, R. L., Demirel, M. C.,  
Keskin, O. and  Bahar, I. (2001) {\it Biophys. J.} {\bf 80}, 505-515. 

\bibitem{ela3}
Takano, M., Higo, J., Nakamura, H. K. and Sasai, M. (2004) 
{\it Nat. Comput.} {\bf 3}, 377-393. 

\bibitem{pdb} RCSB PROTEIN DATA BANK, \\
http://www.rcsb.org/pdb/Welcome.do.   

\bibitem{1kk7}
Himmel, D.M.,  Gourinath, S.,  Reshetnikova, L.,  Shen, Y.,  
Szent-Gyorgyi, A.G. and  Cohen, C. (2002) 
{\it Proc. Natl. Acad. Sci. USA.}  {\bf 99}, 12645-12650. 

\bibitem{austin}
Austin, R. H., Einstein, L., Frauenfelder, H. and Gunsalus, I. C. 
(1975) {\it Biochemistry} {\bf 14}, 5355-5373. 

\bibitem{yana}
Ishijima, A., Kojima, H., Funatsu, T., Tokunaga, M., Higuchi, H., 
Tanaka, H. and Yanagida, T. (1998) {\it Cell} {\bf 92}, 161-171. 

\bibitem{yana2}
Kitamura, K., Tokunaga, M., Iwane, A. H. and Yanagida, T. (1999) 
{\it Nature} {\bf 397}, 129-134. 

\bibitem{yana3}
Iwaki, M., Tanaka, H., Iwane, A. H., Katayama, E., Ikebe, M. 
and Yanagida, T. (2006) {\it Biophys. J.} {\bf 90}, 3643-3652. 

\bibitem{togashi}
Togashi, Y. and Mikhailov., A. S. (2007)  
{\it Proc. Natl. Acad. Sci. USA.}  {\bf 104}, 8697-8702.  

\bibitem{iso}
 Houdusse, A. and Sweeney, H. L. (2001) {\it Curr. Opin. Struct. 
Biol.} {\bf 11}, 182-194. 


\bibitem{fold}
Micheletti, C., Lattanzi, G. and Maritan, A. (2002) {\it  
J. Mol. Biol.}  {\bf 321}, 909-921.




\end{thebibliography}
\end{document}